\documentstyle[aps]{revtex}

\def\ua{\uparrow}
\def\da{\downarrow}
\input{psfig.sty}

\begin{document}
\title{\bf Phase diagram of three-leg ladders at strong coupling along the
rungs}
\author{Maxim Yu. Kagan$^{*}$, Stephan Haas$^{**}$ and T.M.Rice$^{***}$}
\address{
$^{*}$ P.L.Kapitza Institute for Physical Problems, Kosygin street,
Moscow 117334, Russia}
\address{
$^{**}$ Department of Physics and Astronomy, University of Southern California,
Los Angeles, CA 90089-0484, USA
}
\address{$^{***}$Theoretische Physik, ETH-H\"{o}nggerberg, CH-8093
Z\"{u}rich, Switzerland}
\maketitle
\begin{abstract}
A phase diagram of the $t-J$ three-leg ladder as a function of hole dopping is
derived in the limit where the coupling parameters along the rungs, $t_{\perp}$
and $J_{\perp}$, are taken to be much larger than those along the legs,
$t_{||}$ and $J_{||}$ At large exchange coupling along the rungs,
$J_{\perp}/t_{\perp}> \frac{3}{\sqrt{2}}$, there is a transition from a
low-dopping Luttinger liquid phase into a Luther-Emery liquid at a critical
hole concentration $n_{crit}\approx 1/3$. In the opposite case,
$J_{\perp}/t_{\perp}< \frac{3}{\sqrt{2}}$, there as a sequence of three
Luttinger liquid phases (LLI, LLII and LLIII) as a function of hole dopping.
\end{abstract} 
\vspace{.8cm}

\section{Introduction} The recent experimental success in synthesizing
quasi-one-dimensional ladder materials with a mobile charge carriers has raised
an increased interest in the theoretical understanding of their rich phase
diagram \cite{1}, \cite{2}. In previous studies of microscopic models on
various ladder geometries, a competition between superconducting, phase
separation and density wave instabilities has been observed \cite{3}, \cite{4}.
In particular, it was seen that ladders with even and odd number of legs have
quite distinct generic features, such as the presence of a spin gap at
half-filling in even leg system and its absence for odd-leg ladders. When
examining this half-filled case it was realized that inter-band scattering
processes are relevant, and that this is the reason why approximations based on
strong coupling anisotropies, such as expansions in $J_{||}/J_{\perp}$, give a
correct physical picture which extends beyond the isotropic regime.
Furthermore, most ladder materials show coupling anisotropies within the ladder
complex, {\it e.g.} a recent structural analysis of the vanadate ladder
$NaVa_2O_5$ suggests a strong rung-coupling anisotropy of
$J_{\perp}/J_{||}\approx4$ \cite{5}.
\vspace{34pt}
\begin{figure}
\centerline{   \hspace{7mm}\psfig{figure=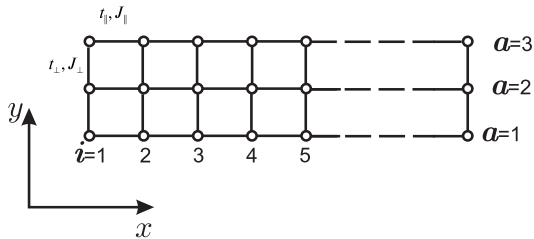,width=7cm}}
\vspace{3mm}
\caption{
The $t-J$ ladder with three legs and L rungs. The couplings along the
legs are $t_{||}$ and $J_{||}$, and those along the rungs $t_{\perp}$ and
$J_{\perp}$.
}
\label{Fig1}
\end{figure}

In this paper we further explore the strong rung-coupling limit of the $t-J$
model in the presence of mobile holes on a three-leg ladder (Fig.1).

By comparing the results in this analytically tractable limit with numerical
diagonalizations, we will see how far this analysis can be extended towards the
regime of isotropic coupling parameters.

We arrive at a phase diagram valid in the limit $J_{\perp}, t_{\perp}>>J_{||},
t_{||}$. At $J_{\perp}<\frac{3}{\sqrt{2}}t_{\perp}$ it contains three various
Luttinger Liquid (LL) phases of different nature, depending on the
concentration of holes. For $J_{\perp}>\frac{3}{\sqrt{2}}t_{\perp}$ at low hole
doping, there is also a LL phase. However beyond a critical hole concentration
a spin gap opens up, and a transition occurs to a Luttinger-Emery liquid (LEL)
with an effective hole-hole attraction, as it has already been observed in
previous studies of the isotropic case ($J_{\perp}=J_{||}$, $t_{\perp}=t_{||}$)
\cite{6}, \cite{7}.

\section{Single-rung states}

The Hamiltonian of the anisotropic $t-J$ model on a three-leg ladder is given
by

\begin{eqnarray}
H= -t_{||} \sum_{i,\sigma} \sum_{a=1}^{3} P(c_{i,a,\sigma}^\dagger
c_{i+1,a,\sigma}+H.c)P -
t_{\perp} \sum_{i,\sigma} \sum_{a=1}^{2} P(c_{i,a,\sigma}^\dagger
c_{i,a+1,\sigma}+H.c)P+ \\
+J_{||} \sum_{i} \sum_{a=1}^{3} \left( \hbox{\bf S}_{i,a} \hbox{\bf S}_{i+1,a} -
\frac{1}{4} n_{i,a}n_{i+1,a} \right) +
\left( J_{\perp} \sum_{i} \sum_{a=1}^{2} \hbox{\bf S}_{i,a} \hbox{\bf
S}_{i,a+1} - \frac{1}{4} n_{i,a}n_{i,a+1} \right)   \nonumber
\end{eqnarray}
where
{\it i} runs over L rungs, $\sigma(= \ua \da)$ and $a$ are spin and leg indices. The
first two terms are the kinetic energy (P is projection operator which
prohibits double occupancy) and the last two exchange couplings $J_{||}
(J_{\perp})$ act along the legs (rungs).

Let us start by discussing the low-energy states of H on a single 3-site rung
with 0,1 and 2 holes in the limit where $J_{||}=t_{||}=0$. They are listed in
Tab.1. In this limit, the exact ground state of the whole ladder is simply a
product of these rung states. An important symmetry present in the 3-leg ladder
is its reflection parity about the center leg, $R$. Along with the total
spin quantum number, $S$, and its projection, $S_z$, $R$ characterized
the symmetry of the ground state vector.

\begin{center}
Table 1. Ground state energies and vectors for \\ a
3-site rung with 0, 1, 2, and 3 holes.
\end{center}
\begin{center}
\begin{tabular}{|c|c|c|}
  \hline
$n$ & $E_{n}$ & ground state eigenvector \\ [6pt]
  \hline
$0$ & $-\frac{3}{2}J_{\perp}$ & $\hbox{~}\frac{1}{\sqrt{6}}[c^{+}_{i,1,\ua}
c^{+}_{i,2,\ua} c^{+}_{i,3,\da} -2 c^{+}_{i,1,\ua} c^{+}_{i,2,\da}
c^{+}_{i,3,\ua}+ c^{+}_{i,1,\da} c^{+}_{i,2,\ua} c^{+}_{i,3,\ua}]  |0,0,0>
 \hbox{~}$ \\ [6pt] \hline $1$ &$\hbox{~}-\frac{4t_{\perp
}^{2}}{\sqrt{J_{\perp }^{2}+8t_{\perp }^{2}} -J_{\perp}}=-\frac{2t_{\perp
   }}{\alpha _{1}}\hbox{~}$ & $\frac{1}{\sqrt{4+2\alpha
_{1}^{2}}}[c^{+}_{i,1,\ua} c^{+}_{i,2,\da} - c^{+}_{i,1,\da} c^{+}_{i,2,\ua}
+\alpha_{1} c^{+}_{i,1,\ua}  c^{+}_{i,3,\da} -\alpha_{1} c^{+}_{i,1,\da}
c^{+}_{i,3,\ua} + $\\
&&$+c^{+}_{i,2,\ua} c^{+}_{i,3,\da}  - c^{+}_{i,2,\da}
c^{+}_{i,3,\ua} ] |0,0,0>$ \\ [6pt]
 \hline
$2$ & $-\sqrt{2}t_{\perp }$ &
$\frac{1}{2}[c^{+}_{i,1,\ua} +\sqrt{2} c^{+}_{i,2,\ua}+ c^{+}_{i,3,\ua} ]
|0,0,0>$ \\ [6pt]
 \hline  $3$ & $0$ & $|000>$ \\ [6pt] \hline  \end{tabular}
\end{center}

At half filling (0 holes) the ground state is two-fold degenerate
(S=$\frac{1}{2}$; S$_z = \pm \frac{1}{2}$), and it has odd parity with
respect to reflection about the center leg, $R=-1$. As discussed  previously
[6,8], this state behaves as an effective spin -1/2 rung spin, and an
inter-rung magnetic coupling $J_{||}$ between such states introduces the
low-energy behavior of a Heisennberg AFM spin -1/2 chain.

Note that first excited state here is also ($S=\frac{1}{2}$;
$S_z=\pm\frac{1}{2}$) doublet, but it has even reflection parity, $R=1$, and
thus is in a non-bonding configuration. A second excited state corresponds
already to $S=\frac{3}{2}$ and is irrelevant for our considerations.

The ground state with one hole on a 3-site rung is a singlet ($S=S_z=0$) with
even parity, $R=1$. Subsequently, we will consider separately the two regimes
of " strong coupling" $J_{\perp} >> t_{\perp}$ and "weak coupling"
$J_{\perp}<<t_{\perp}$.  In strong coupling case $E_1 \approx - J_{\perp} - 2
{t_{\perp}^2 \over{J_{\perp}}}$, while in weak coupling case  $E_1 \approx -
\frac{J_{\perp}}{2} - \sqrt{2} t_{\perp}$.  It will be shown later that
hole-hole pairing on a rung (leading to LEL) occurs naturally in the strong
coupling limit, while it is absent at weak couplings.

The first excited state corresponds to a nonbonding singlet with   $R=-1$,
while a second excited state  -- to an antibonding singlet with $R=1$.

For 2 holes on a 3--site rung, the ground state  is a ($S=\frac{1}{2}$ ; $S_z=
\pm \frac{1}{2}$) doublet with $R=1$. The first excited state is non-bonding
with $R=-1$, while a second excited state is antibonding with $R=1$ again.

So let us compare now  in  the limit of almost independent rungs
($J_{||}=t_{||}=0$) the ground state energies for different configurations.
As a result we obtain:

\begin{enumerate}
\item for hole concentration $0<n<\frac{1}{3}$ a minimal energy
$min E=E_a = (1-3n)E_0 L +3n E_1 L$ --corresponds to a mixture of rungs with
one hole and without holes (Fig.2)
\begin{figure}
\centerline{\hspace{7mm} \psfig{file=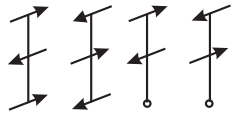,width=3cm}}
\caption{
  Configuration a). Rungs with one hole in the surrounding of rungs without
holes
}
\label{Fig2}
\end{figure}
\item for $\frac{1}{3}<n< \frac{2}{3}$  -- two configurations are possible:
\begin{equation}
\left\{
   \begin{tabular}{l}
   $ E_b = (2-3n)E_1L +3nE_2L$ \\
    $E_c = \frac{3}{2}(1-n)E_1L+(\frac{3}{2}n-\frac{1}{2})E_3L = \frac{3}{2}
    (1-n)E_1 L $
    \end{tabular}
  \right.    \nonumber
\end{equation}
\begin{figure}
\centerline{\hspace{7mm} \psfig{file=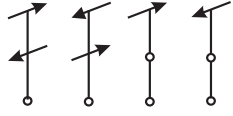,width=3cm}}
\caption{
  Configuration b). Rungs with one hole in the surrounding of rungs with two
holes.
}
\label{Fig3}
\end{figure}
\begin{figure}
\centerline{\hspace{7mm} \psfig{file=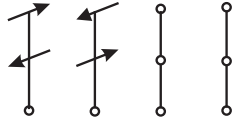,width=3cm}}
\caption{
  Configuration c). Rungs with one hole in the surrounding of rungs with three
holes.
}
\label{Fig4}
\end{figure}
Configuration b) is given by Fig.2 and corresponds to the mixture of rungs with
one and two holes, while configuration c) is given by Fig.4 and corresponds to
the mixture of one and three holes.
\item Finally for $\frac{2}{3}<n<1$  there are two possibilities again:
$E_c= \frac{3}{2}(1-n)E_1 L $ -- is given by Fig.4 again while
$E_d = 3(1-n)E_2 L$  --- corresponds to a mixture of rungs with 3 and 2 holes
(Fig.5 )
\begin{figure}
\centerline{\hspace{7mm} \psfig{file=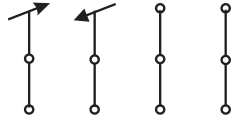,width=3cm}}
\caption{
  Configuration d). Rungs with two holes in the surrounding of rungs with three
holes.
}
\label{Fig5}
\end{figure}

Comparison of $E_d$ and $E_c$ yields:
$E_2 = \frac{1}{2} E_1$  and  hence:
$-2\sqrt{2}t_{\perp} = - \frac{4 t_{\perp}^2}{\sqrt{J_{\perp}^2+8 t_{\perp}^2} -
J_{\perp}}$

As a result:
\begin{equation}
(\frac{J_{\perp}}{t_{\perp}})_{crit} = \frac{3}{\sqrt{2}}
\end{equation}

For $J_{\perp}> J_{\perp crit} $  in all the region $\frac{1}{3}<n<1$
configuration c) [3+1] holes is realized.
For$J_{\perp}<J_{\perp crit}$ in the region $\frac{1}{3}<n<\frac{2}{3}$
configuration b) [1+2] holes is realized, while for $\frac{2}{3}<n<1$
configuration d) [3+2] holes is more beneficial. (see Fig.6)
\begin{figure}
\centerline{\hspace{7mm} \psfig{file=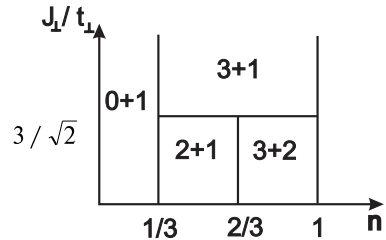,width=5cm}}
\caption{
  Phase diagram for different ground state configurations as a function of
  $J_{\perp}/t_{\perp}$
}
\label{Fig6}
\end{figure}

\end{enumerate}

\section{Kinetic energy of rungs delocalization}

Let us "switch on" the next approximation and calculate kinetic energy of rungs
delocalization.

To be more specific for [0+1] phase we need to calculate a matrix element which
interchange the rung with 0 holes and the rung with 1 hole. (Fig. 7)
\begin{figure}
\centerline{\hspace{7mm} \psfig{file=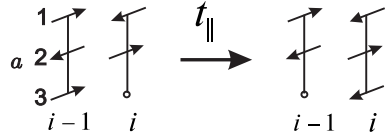,width=5cm}}
\caption{
Interchange of the rung with 0 holes and 1 hole in first order in
$t_{||}$.
}
\label{Fig7}
\end{figure}

On electronic language we calculate:
$$
  -t_{||}\sum_{a,\sigma}c^{+}_{i,a,\sigma} c_{i-1,a,\sigma} \left| \Psi_i(2 el)
  \Psi_{i-1}(3 el)\right> \equiv -t_{||}\sum_{a,\sigma}\left|c^{+}_{i,a,\sigma}
 \Psi_i(2 el)\right> \left| c_{i-1,a,\sigma} \Psi_{i-1}(3 el)\right>,
$$
where $c^{+}_{i,a,\sigma}$, $c_{i-1,a,\sigma}$ are electronic operators;
$$
\Psi_{i}(2 el) \equiv \Psi_{i}(1 h); \hbox{~~~} \Psi_{i-1}(3 el) \equiv
\Psi_{i-1}(0 h) \hbox{~~ (see Table 1)}
$$

Note that large scales $J_{\perp}$ and $t_{\perp}$ fix the global
configuration, {\it i.e.} for given concentration of holes both
the number of rungs with 2 electrons and
the number of rungs with 3 electrons do not change.

That is why:
$$
  -t_{||}\sum_{a,\sigma}\left|c^{+}_{i,a,\sigma}
 \Psi_i(2 el)\right> \left| c_{i-1,a,\sigma} \Psi_{i-1}(3 el)\right>=-t_{eff}
 \Psi_i(3 el) \Psi_{i-1} (2 el).  $$

 The calculation of the $t_{eff}$ yields:
\begin{equation}
  t_{eff}=\frac{3t_{||}}{4} \frac{\alpha_2}{\alpha_2-\alpha_1},
  \end{equation}
where
$\alpha_{1,2}=-\frac{J_{\perp}\pm\sqrt{J_{\perp}^2+8t_{\perp}^2}}{2t_{\perp}}$ -
coefficients which enter, respectively, in the eigenvectors of a ground state
(see eq.(3)) and an antibonding state for 2 electrons on the rung.

Note that to calculate $t_{eff}$ we used anticommutation relations for
fermionic operators $c^+$ and $c$ together with a condition which prohibits a
double occupancy.

In the limiting cases expression (4) reads:
$$
t_{eff} = \left\{
  \begin{array}{l}
    \frac{3t_{||}}{4}\left[ 1+O\left( \frac{J_{\perp
    crit}^2}{J_{\perp}^2}\right) \right] \hbox{~~~for~~~} J_{\perp} >> J_{\perp
    crit}, \\ [6pt]
  \frac{3t_{||}}{8}\left[ 1+ \frac{3 J_{\perp}}{4 J_{\perp
    crit}} \right] \hbox{~~~~~~for~~~} J_{\perp} << J_{\perp crit} .
     \end{array}
\right.
$$

In the language of effective operators:
\begin{equation}
  \hat{H}_{kin}=-t_{eff} \sum_{i\sigma} a^{+}_{i,\sigma}a_{i-1,\sigma}.
  \end{equation}

An operator $a^{+}_{i,\sigma}$ corresponds to the creation on the site $i$ of a
rung with 3 electrons and simultaneous destruction on the same site of a rung
with 2 electrons. So, we could represent $a^{+}_{i,\sigma}$ in the following
form
\begin{equation}
  a^{+}_{i,\sigma}=f^{+}_{i,\sigma} b_i
  \end{equation}

Here an operator $f^{+}_{i,\sigma}$ creates the rung with 3 electrons and total
spin $S=1/2$. Hence it has a fermionic nature.

An operator $b_i$ destroys 2 electrons (an electronic singlet with total spin
$S=0$) and hence has a bosonic nature.

Of course, 2 rungs could not occupy the same place. It means that they are
subject of infinitely strong Hubbard repulsion:
\begin{equation}
  U_{\infty}\sum_{1\sigma} \rho_{i,\sigma} \rho_{i,-\sigma},
  \end{equation}
where
$\rho_{i\sigma}=a^{+}_{i,\sigma}a_{i,\sigma}
= (f^{+}_{i,\sigma} f_{i,\sigma})(b^{+}_{i,\sigma} b_{i,\sigma})$.
So, in a state [0$+$1] holes a system is described by effective Hamiltonian:
\begin{equation}
  \hat{H}_{0+1}=E_{0+1}-t_{eff}\sum_{i\sigma}a^{+}_{i,\sigma}a_{i,\sigma}+
  U_{\infty}\sum_{i} {\rho_{i,\sigma}} {\rho_{i,-\sigma}}.
  \end{equation}

It is 1D fermionic Hubbard model with repulsion. We know that it belongs to the
universality class of Luttinger liquid \cite{9}.

Note that a more detailed analysis in case of $J_{\perp}>>t_{\perp}$ shows that
at densities $n^*
 {\ \lower-1.2pt\vbox{\hbox{\rlap{$
        >
        $}\lower5pt\vbox{\hbox{$\sim$}}}}\ }
\left( \frac{t_{\perp}}{J_{\perp}} \right)^{1/2}$ (which,
in principle, could be smaller than $1/3$) we will have a two-band degenerate
Hubbard model instead of a one band model. However, this situation will also
fall in the universality class of Luttinger liquid I.

Let us consider now a state [2+1]. Here:
$$
t_{eff}=\left<\Psi_i(2 el) \Psi_{i-1}(1 el)\left| -t_{||} c^{+}_{i,a,\sigma}
c_{i-1,a,\sigma} \right| \Psi_i(1 el) \Psi_{i-1}(2 el) \right>
$$

Direct calculation of $t_{eff}$ yields:
\begin{equation}
  t_{eff}=\frac{t_{||}}{2} \frac{1}{(\alpha_2-\alpha_1)} \left[ \left(
  1+\frac{\alpha_1}{2\sqrt{2}} \right) \alpha_2 -\frac{(\sqrt{2}+\alpha_1)}{2}
  \right].
  \end{equation}

For the limiting cases:
$$
t_{eff} = \left\{
  \begin{array}{l}
    \frac{t_{||}}{2}\left[ 1+ \frac{2J_{\perp crit}}{3J_{\perp}}
    \right] \hbox{~~~for~~~} J_{\perp} >> J_{\perp crit}, \\ [6pt]
    \frac{5t_{||}}{8}\left[ 1+ \frac{3 J_{\perp}}{20 J_{\perp crit}}
    \right] \hbox{~~for~~~} J_{\perp} << J_{\perp crit} .
     \end{array}
\right.
$$

As a result:
$$
\hat{H}_{kin}=-t_{eff}\sum d^{+}_{i,\sigma} d_{i-1,\sigma}.
$$

An effective operator $d^{+}_{i,\sigma}=f^{+}_{i,\sigma}b_{i}$ has a fermionic
nature again (see Fig.8). It creates a rung with 1 electron and $S=1/2$
($f^{+}_{i,\sigma}$), and simultaneously destroys a rung with 2 electrons and
$S=0$ ($b_i$). Finally:
\begin{equation}
  \hat{H}_{2+1}=E_{2+1}-t_{eff}\sum d^{+}_{i,\sigma}d_{i-1,\sigma} +U_{\infty}
  \sum \tilde{\rho}_{i,\sigma} \tilde{\rho}_{i,-\sigma},
  \end{equation}
where
$$
\tilde{\rho}_{i,\sigma}=d^{+}_{i,\sigma}d_{i,\sigma}
$$
\begin{figure}
\centerline{\hspace{7mm} \psfig{file=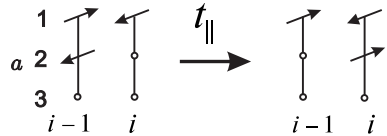,width=5cm}}
\caption{
  Interchange of the rung with 1 electron and the rung with 2 electrons due to
  $t_{||}$.
} \label{Fig8}
\end{figure}

We again derive a 1D fermionic Hubbard model with repulsion. So a state [2+1]
corresponds to LLII which describes a motion of a rung with 1 electron in the
surrounding of rungs with 2 electrons.

Now let us proceed to the case [3+2] (Fig.9). Here
$$
t_{eff}=\left<\Psi_i(1 el) \left| -t_{||} c^{+}_{i,a,\sigma}
c_{i-1,a,\sigma} \right| \Psi_{i-1}(1 el) \right>
$$
\begin{figure}
\centerline{\hspace{7mm} \psfig{file=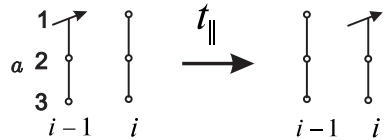,width=5cm}}
\caption{
  Interchange of the rung with 1 electron and an empty rung due to $t_{||}$.
}
\label{Fig9}
\end{figure}

It is easy to derive that $t_{eff} \equiv t$ because this problem is equivalent
to the motion of an electron in the empty space.

Hence:
$$
\hat{H}_{kin}=-t_{eff}\sum f^{+}_{i,\sigma} f_{i-1,\sigma},
$$
where $f^{+}_{i,\sigma} \equiv \Psi_i (1 el)$ - has a fermionic nature again.
(It creates a rung with 1 electron and destroys an empty rung). As a result:
\begin{equation}
  \hat{H}_{3+2}=E_{3+2}-t_{||}\sum f^{+}_{i,\sigma} f_{i-1,-\sigma} +U_{\infty}
  \sum \hat{x}_{i,\sigma} \hat{x}_{i,-\sigma},
  \end{equation}
where
$
{\hat{x}_{i,\sigma}}=f^{+}_{i,\sigma}f_{i,\sigma}
$
is a density of rungs with one electron.

Hamiltonian (11) again corresponds to 1D Hubbard model with repulsion. So, it
describes LLIII where the rungs with 1 electron move in the surrounding of
rungs without electrons.

A state [3+1] has quite a different nature.

Here $t_{eff} \sim \frac{t_{||}^2}{|E_1-2E_2|}$ (see Fig.10).
\begin{figure}
\centerline{\hspace{7mm} \psfig{file=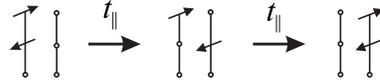,width=5cm}}
\caption{
  Interchange of the rung with 2 electrons and an empty rung in second order
  of perturbation theory in $t_{||}$.
}
\label{Fig10}
\end{figure}

Since a phase [3+1] is realized only for $J_{\perp}>J_{\perp crit}$: $t_{eff}
\sim t_{||}^2/J_{\perp}$. Kinetic energy in this state describes a motion of
the rung with 2 electrons in the surrounding of empty rungs. Hence
\begin{equation}
\hat{H}_{kin}=-t_{eff}\sum b^{+}_{i} b_{i-1},
\end{equation}
where $b^{+}_{i} \equiv \Psi_i (2 el)$ is a bosonic operator which creates a
rung with 2 electrons and destroys an empty rung. Of course, 2 rungs with 2
electrons can not occupy the same place. As a result:
\begin{equation}
  \hat{H}_{3+1}=E_{3+1}-t_{eff}\sum b^{+}_{i} b_{i-1} +U_{\infty}
  \sum {n^i_{bos}} {n^i_{bos}},
  \end{equation}
$
{n^i_{bos}}=b^{+}_{i} b_{i},
$
and now we have 1D Bose-Hubbard with strong repulsion. This model belongs to
universality class of Luther-Emery liquid. It has a spin gap at half filling
and large superconductive fluctuations in a doped case. When we include a boson
rescattering between neighbouring ladders, than a finite $T_C$ arises in the
system \cite{10}.

Finally, the phase diagram reads (Fig.11):
\begin{figure}
\centerline{\hspace{7mm} \psfig{file=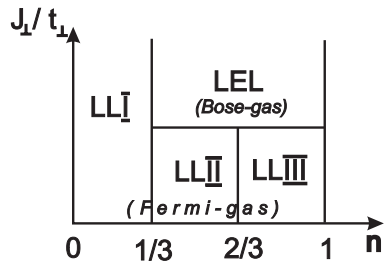,width=5cm}}
\caption{
  The phase diagram of three-leg ladder at strong coupling along the rungs.
}
\label{Fig11}
\end{figure}

\section{The role of AFM exchange along the legs $J_{||}$.}

In a close analogy with a double exchange model for
$J_H S >> zt {\ \lower-1.2pt\vbox{\hbox{\rlap{$
        >
        $}\lower5pt\vbox{\hbox{$\sim$}}}}\ } J_{ff} S^2$:
$$
\hat{H}=-t \sum_{i\sigma}c^+_{i\sigma}c_{i\sigma} - J_H \sum_i {\hbox{\bf S}}_i
{\bf {\sigma}}_i + J_{ff} \sum_i {\hbox{\bf S}}_i {\hbox{\bf S}}_{i-1} -
$$
- a largest scale (FM exchange $J_H$) forms a local onsite state with
$S_{tot}=S+1/2$ and then effectively drops out of the model. Low energy physics
(including phase separation on FM and AFM regions) is governed solely by smaller
parameters $t$ and $J_{ff}$. Absolutely the same scenario is realized in our
model. The largest parameters $J_{\perp}$ and $t$ form the stable
configurations LLI, LLII, LLIII, and LEL and after that effectively drop out of
the model. Low-energy physics is governed solely by $J_{||}$ and $t_{||}$.

Then, by analogy with FM-polarons formation in double exchange model, we could
have in our case either a diluted configuration (Fig.12) or a phase-separated
state (clusterization) (Fig.13). The clusterized phase was found in \cite{7} in
numerical study of an isotropic regime $t_{||}=t_{\perp}$, $J_{||}=J_{\perp}$;
$J=0.35t$.
\begin{figure}
\centerline{\hspace{7mm} \psfig{file=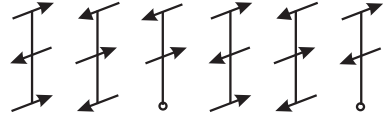,width=5cm}}
\caption{
A diluted configuration in LLI
}
\label{Fig12}
\end{figure}
\begin{figure}
\centerline{\hspace{7mm} \psfig{file=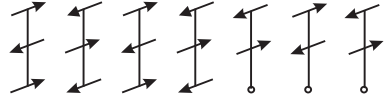,width=5cm}}
\caption{
 Clusterization in LLI
}
\label{Fig13}
\end{figure}

A diluted phase is more beneficial with respect to kinetic energy $\sim
t_{||}$, while a clusterized phase is more beneficial with respect to magnetic
exchange energy between the rungs $\sim J_{||}$. So, in the case of small
$J_{||} {\ \lower-1.2pt\vbox{\hbox{\rlap{$
        <
        $}\lower5pt\vbox{\hbox{$\sim$}}}}\ } t_{||}$
a diluted phase is more beneficial. The energy of this state in case of LLI
reads:
\begin{equation}
  \tilde{E}_{0+1}=(1-3n)(E_0+E_{00}^J)L + 3n(E_1+E^{kin}_{01}-E^J_{00})L.
  \end{equation}
Here $E_{01}^{kin}=-2t_{eff}$ (where $t_{eff}$ is given by (9)) - corresponds
to delocalization energy when rungs with one hole occupy the bottom of the
band.
$$
E_{00}^J=\left< \Psi_i^+(3 el) \Psi_{i-1}^-(3 el) \left| J_{||} \sum_a
({\hbox{\bf S}}_{i,a} {\hbox{\bf S}}_{i-1,a} -\frac{1}{4} n_{i,a} n_{i-1,a})
\right| \Psi_i^+(3 el) \Psi_{i-1}^+(3 el) \right> =-J_{||} $$ -corresponds to a
chessboard AFM configuration of rungs with 0 holes (see Fig.14).
\begin{figure}
\centerline{\hspace{7mm} \psfig{file=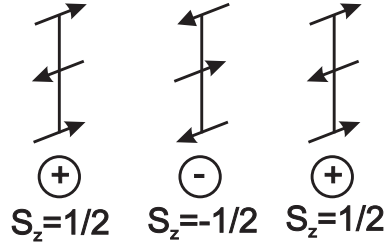,width=5cm}}
\caption{
A chessboard AFM configuration of rungs with 0 holes.
}
\label{Fig14}
\end{figure}

Effective Hamiltonian for LLI reads:
\begin{equation}
\hat{H}=\tilde{E}_{0+1}-t_{eff} \sum_i a^+_{i,\sigma} a_{i-1,\sigma}
+U_{\infty} \sum_i {\rho}_{i,\sigma} {\rho}_{i-1,\sigma} + J_{||}
\sum_i {\hbox{\bf S}}_i {\hbox{\bf S}}_{i-1},
\end{equation}
where ${\hbox{\bf S}}_i$ is a rung spin on site $i$.

So, a final effective $H$ is given by $t_{eff}-J_{||}$ model.

For $J_{||} \leq 2 t_{eff}$ it still belongs to the universality class of LL.
There is no spin gap at half-filling. The basic instability for moderate doping
is towards SDW-formation \cite{11}.

For $J_{||}> 2 t_{eff}$, in total analogy with 2D $t-J$ model, a bound state
appears and model becomes unstable towards superconductivity. So, for small
$J_{||} {\ \lower-1.2pt\vbox{\hbox{\rlap{$
        <
        $}\lower5pt\vbox{\hbox{$\sim$}}}}\ } t_{eff}$
its role is just:
\begin{enumerate}
  \item to form small energy corrections,
  \item to change a little bit phase boundaries on a phase diagram:
  $$
  n_{crit}=\frac{1}{3} \rightarrow \tilde{n}_{crit} = \frac{1}{3} \left[ 1-\eta
  \frac{J_{||}}{t_{||}} \right]^{1/2},
  $$
   $\eta$ is a numerical coefficient,
  \item to introduce in the model effective AFM attraction between neighbouring
    rungs.
\end{enumerate}

As a result the gross features of phase diagram on Fig.11 are conserved.

\section{Discussion}

Now let us proceed from the calculations to qualitative arguments. In isotropic
case, besides a tendency towards clusterization \cite{7}, there is a tendency
towards coexistence of LL and LEL. (Fermi-Bose liquid scenario restated recently
in \cite{12}). In strong coupling regime there is a gap between the bottoms of
Fermi-gas band and a Bose-gas band. The gap is of the order of
$t_{\perp}-2t_{||}$.  In an isotropic regime we could overcome this gap due to
an increase of $t_{||}$ $(t_{||}=t_{\perp})$ and obtain a picture of Fig.15:
\begin{figure}
\centerline{\hspace{7mm} \psfig{file=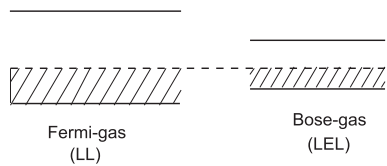,width=5cm}}
\caption{
  Fermi-Bose liquid scenario restated recently in [12].
}
\label{Fig15}
\end{figure}

\section{Conclusion}

In conclusion let us emphasize that
\begin{enumerate}
  \item 3-leg ladder possesses both properties of 2-leg ladders and 1D doped
  spin chains.  \item There is a hope that when we increase a number of legs
  the difference between odd and even numbers will become smaller. In favour of
  this assumption is a fact that in 2n-leg ladders spin-gap scales as:  $$
  \Delta_{2n} \sim \Delta_2 /(2^n) \sim J_{\perp}/2^n \rightarrow 0
   \hbox{~~~for~~~} n \rightarrow \infty.
   $$
  \item HTSC-materials have both properties of two and three leg ladders.
   \end{enumerate}

At
$n {\ \lower-1.2pt\vbox{\hbox{\rlap{$
        <
        $}\lower5pt\vbox{\hbox{$\sim$}}}}\ } (2\div 4)\%$
(AFM region) there is no spin gap in HTSC materials as in 3-leg ladders for
$J_{\perp}<t_{\perp}$.

However at
$(2 \div 4)\% {\ \lower-1.2pt\vbox{\hbox{\rlap{$
        <
        $}\lower5pt\vbox{\hbox{$\sim$}}}}\ } n$
${\ \lower-1.2pt\vbox{\hbox{\rlap{$
        <
        $}\lower5pt\vbox{\hbox{$\sim$}}}}\ } 20\%$
there is a spin pseudogap in HTSC materials in analogy with 2-leg ladders. It
means that HTSC has a long prehistory when we go from half-filling ($n=0$) to
an optimal doping values ($n=0.15$).

The authors are grateful to I.A.Fomin for valuable comments.
M.Yu.K. acknowledges the hospitality of Theoretical Department in ETH Z\"{u}rich
where this work was started, and is also grateful to the President Eltsin grant
N 98-15-96942 for financial support.

\end{document}